\documentclass[twocolumn,a4paper,superscriptaddress,prb]{revtex4-1}

\usepackage{graphicx, epsfig}
\usepackage{amssymb,amsmath}
\usepackage{color}

\usepackage{times}

\def \hf{\tfrac{1}{2}}

\newcommand{\bra}[1]{\langle\left.{#1}\right|}
\newcommand{\ket}[1]{\left|{#1}\right.\rangle}
\newcommand{\xpct}[1]{\langle{#1}\rangle}    

\newcommand{\TK}{T_{\rm K}}

\global\long\def\av#1{\left\langle #1 \right\rangle }

\global\long\def\abs#1{\left|#1\right|}

\begin{document}

\title{Single-impurity Kondo physics at extreme particle-hole asymmetry}

\author{Yulia E.~Shchadilova}
\affiliation{Max Planck Institute for the Physics of Complex Systems, N\"othnitzer Str.~38, 01187
Dresden, Germany}

\affiliation{A.~M.~Prokhorov General Physics Institute, Russian Academy of Sciences, Vavilov str.~38, 119991 Moscow, Russia}

\affiliation{Russian Quantum Center, Novaya str.~100, 143025 Skolkovo, Moscow region, Russia}

\author{Matthias Vojta}
\affiliation{Institut f\"ur Theoretische Physik, Technische Universit\"at Dresden, 01062 Dresden,
Germany}

\author{Masudul Haque}
\affiliation{Max Planck Institute for the Physics of Complex Systems, N\"othnitzer Str.~38, 01187
Dresden, Germany}

\begin{abstract}

We study the fate of the Kondo effect with one-dimensional conduction baths at very low densities,
such that the system explores the bottom of the conduction band.
This can involve either finite low densities, or a small number of fixed conduction electrons in a
large system, i.e., the limit of large bath sizes can be taken with either fixed small density or
with fixed number.
We characterize the Kondo physics for such systems through the energy gain due to Kondo coupling,
which is a general analog of the Kondo temperature scale, and through real-space profiles of
densities and spin-spin correlation functions.

\end{abstract}

\date{\today}



\maketitle


\section{Introduction}

The single-impurity Kondo model \cite{kondo, hewson} has played a crucial role in the study of
correlated electron and mesoscopic physics for several decades. Central to Kondo physics is the
competition between itinerancy of conduction electrons and magnetic coupling to an immobile
impurity, which leads to many-body screening of the impurity moment at low temperatures. The original
setting \cite{kondo} involves a conduction bath in the thermodynamic limit at finite filling and a
Kondo coupling much smaller than the conduction bandwidth or Fermi energy. As a result, the
scattering processes responsible for screening are restricted to the region around the Fermi
surface where the dispersion can be considered linear and continuous. With the realization of
experimental setups where many-body phenomena can be explored in novel confined geometries, some
attention has been paid to situations such as the ``Kondo box'' where the conduction bath is small
enough for the bath spectrum to be discrete. \cite{ThimmKrohavonDelft_PRL99,
KaulUllmoChandrasekharanBaranger_EPL05,HandKrohaMonien_PRL06,Booth-et-al_PRL05,KaulBaranger_PRB09,
KaulZarandChandrasekharanUllmoBaranger_PRL06} Further generalizations of the Kondo problem involve
non-metallic host systems like superconductors\cite{zitt70,sakaisc,withoff,cassa} or
semimetals.\cite{baskaran07,wehling10,dellanna10,epl,uchoa11b,fvrop}

In this work, we examine a situation that differs from the original context in that the number or
density of mobile carriers is extremely small, but the spatial size of the conduction bath is not
necessarily small. Here we will treat a small number or density of {\em electrons}; these situations
are equivalent to those with a small number or density of {\em holes}.

\begin{figure}[tb]
\begin{center}
  \includegraphics[width=1.03\columnwidth]{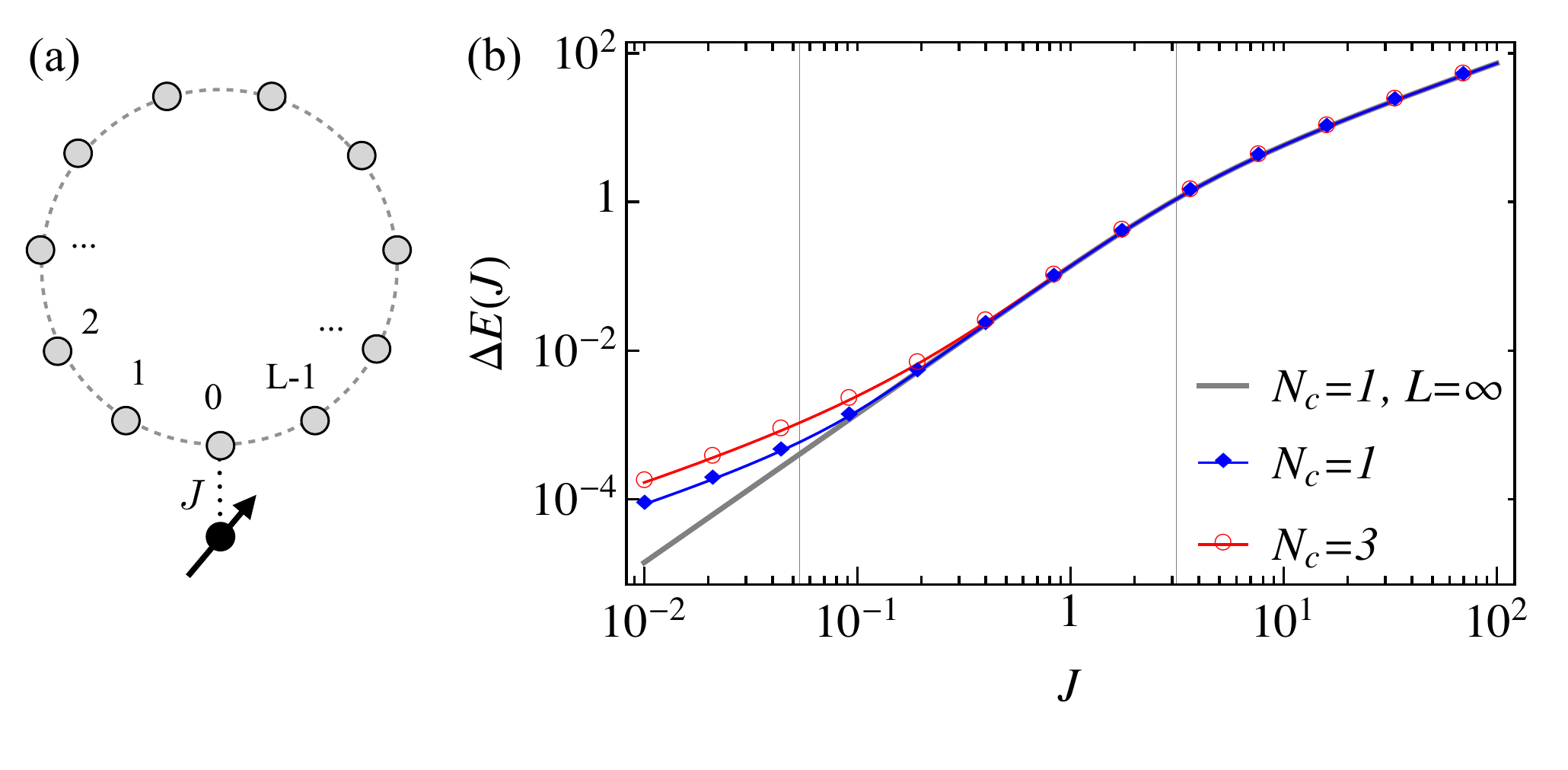}
\end{center}
  \caption{\label{fig:energy} Left: Geometry used in this work.  Right: dependence of the energy
    gain $\Delta E(J)=E_0(0)-E_0(J)$ on Kondo coupling parameter $J$ for a system with $L = 100$
    sites with $N_c=1$ (blue diamonds and line) and $N_c=3$ (red circles and line) conduction
    electrons. The three regions (explained in text) are clearly visible and separated by vertical
    lines.  The curve with no symbols is the analytical $L=\infty$ result for $N_c=1$.  }
\end{figure}

In most of the paper, we shall provide explicit results for $N_c$ mobile fermions (``conduction
electrons'') in a tight-binding chain with periodic boundary conditions, Figure \ref{fig:energy}a. One
site of the lattice is Kondo-coupled to a single spin-$\hf$ ``impurity''.  The Hamiltonian is
\begin{equation} \label{eq:minimalKondoHam}
H ~=~ - \sum_{i,s} \left( c_{i,s}^{\dagger}c_{i+1,s} +
\mathrm{h.c.} \right)
 ~+~  J  \vec{S}_{\mathrm{imp}}\cdot \vec{s_0}
\end{equation}
where $\vec{s_0}=\frac{1}{2}\sum_{s,s'}c_{0,s}^{\dagger}\vec{\sigma}_{ss'}c_{0,s'}$ is the spin on
site $i=0$ ($s$, $s'$ are spin indices), and $i\in[0,L-1]$ is the site index. The Kondo coupling is
antiferromagnetic ($J>0$) and prefers singlet formation. We use the conduction band hopping strength
(quarter the bandwidth) as the unit of energy.  Our regimes of interest are (a) constant number of
bath electrons $N_c$ in a large number of sites, $L\to\infty$, and (b) constant and small density,
i.e., $0<n_c=(N_c/L)\ll\tfrac{1}{2}$ with $L\to\infty$.  The first situation does not correspond to
the usual thermodynamic limit; we will refer to this as the ``ultralow density'' limit. The second
situation is the low-density thermodynamic limit.  While most of our analytical and numerical
results are specific to the case of a one-dimensional (1D) bath, the overall picture is of more
general validity, and we will briefly discuss the modifications occurring for higher-dimensional
geometries.

The case of a few electrons forming the bath (ultralow-density limit, small fixed $N_c$) involves the
competition between itinerancy and antiferromagnetic coupling, which is at the heart of the Kondo
effect, but does not have a true Fermi surface, which is central to the standard analysis of the
single-impurity Kondo problem.  This is thus an important toy model to study the effects of the
above-mentioned competition without the effects of a regular Fermi surface.  The finite but low-density
thermodynamic limit (small fixed $N_c/L$) is closer to the usual situation,\cite{hewson} but explores
the nonlinearity of the lowest part of the band --- there is a Fermi surface but it might not be
possible to linearize around it. The two situations (ultralow density and low density) are thus only
loosely related to each other, and are both different from the usual Kondo setup and from the ``Kondo
box'' (small $L$) situation.

It is quite conceivable that one or both of the regimes we study here might be realized in either
mesoscopic setups, or cold atoms, or both.  In a mesoscopic situation, a low density or low number
of bath electrons might possibly be achieved by appropriately gating the bath.  Small numbers or
densities are completely natural with cold atom experiments, although a cold atom realization of
Kondo physics is not yet available.  (Ref.\ \onlinecite{Kondo_in_cold_atom} proposes such a
realization.)

We characterize single-impurity Kondo physics in the low-density or small-number situation in two ways.
First, we provide results on the energy gained due to the impurity, i.e., the ground state energy
without the impurity coupling subtracted from the ground state energy with the impurity coupling,
${\Delta}E(J)=E_0(0)-E_0(J)$.  This quantity is the analog of the Kondo temperature $\TK$ well-known in
usual Kondo physics,\cite{hewson} and thus clearly an observable of central importance for any type of
Kondo physics.  (We use the notations $\TK$ and ${\Delta}E$ interchangeably; depending on the
definition of $\TK$ they differ by an unimportant constant factor.)  Second, we look at real-space
profiles, in the spirit of recent work examining the ``Kondo cloud''. \cite{SimonAffleck_PRB01,
SimonAffleck_PRL02, SimonAffleck_PRB03, AffleckSimon_PRL01,
  BusserAndaDagotto_PRB10, Simonin_arXiv07, Goth_Assaad, Holzner-etal_PRB09, Borda_PRB07,
  Bergmann_PRB08, BordaGarstKroha_PRB09, Saleur, MitchellBulla_PRB11, SorensenLaflorencieAffleck,
  PereiraLaflorencieAffleckHalperin_PRB08, SougatoBose_kondocloud, BarzykinAffleck_PRB98, ishii,
  SorensenAffleck96} We present results on the conduction electron density,
$n_j=\sum_{\sigma}\xpct{c_{j\sigma}^{\dagger}c_{j\sigma}}$, and the impurity-bath spin correlator,
$\chi_j=\xpct{ \vec{S}_{\mathrm{imp}}\cdot \vec{s_j}}$, as a function of the distance $j$ from the
impurity.  At and near half filling, the shape of the function $\chi_j$ is often described as
characterizing the Kondo cloud, while the conduction electron is not strongly affected by the
magnetic impurity. Away from half-filling, the spin and charge sectors are strongly coupled.  In
fact, in the extreme limit of a single particle, we show that the two profiles, $n_j$ and $\chi_j$,
are identical.  As half-filling is approached, the real-space profies of $n_j$ and $\chi_j$ become
more and more decoupled.  For $N_c=1$, we also describe the spatial structures in terms of the
entanglement entropy between a block including the impurity and the rest of the system.  This is
motivated by recent descriptions of impurity screening clouds using such entanglement entropies.
\cite{SorensenLaflorencieAffleck, SaleurVasseur_PRB13, GhoshRibeiroHaque_2013}

In the original setting for Kondo physics, the coupling $J$ is small compared to the other energy
scales such as the bandwidth or Fermi energy.  Since the regimes considered here are expected to be
relevant to new settings for Kondo physics, we consider $J$ values from $J\ll1$ to $J\gg1$ without
restriction.

Figure \ref{fig:energy} summarizes the behavior of $\Delta{E}(J)$ for small, intermediate and large
$J$, for a fixed number of particles ($N_c=1$ and $N_c=3$) in a large ring with $L\gg1$.  For any
finite $L$, there are three clearly different regions of $J$ values, which we will refer to as regions
A, B, and C from small to large $J$.
Region A (small $J$) is where the Kondo coupling is perturbative. Hence $\Delta{E}(J)$ is linear, with
a coefficient that vanishes as $\sim{L}^{-1}$ at large $L$.
In region C (large $J$), the Kondo coupling $J$ is so strong that the impurity simply binds one fermion to
it in a singlet.  As far as the other electrons are concerned, the impurity connected site ($j=0$) is
blocked and the ring is cut into an open chain of $(L-1)$ sites.  The energy gain is the singlet energy,
$\Delta{E}(J)\approx\frac{3}{4}J$.
Between these two linear-$J$ regions lies the nonperturbative region B.  We will present evidence
that there the behavior is $\Delta{E}(J){\sim}J^2$ in 1D for fixed small $N_c$.   This also applies for
small densities $n_c$ and not too small $J$.  For infinite $L$ the region A disappears and region B extends all the way down to infinitesimal $J$; this is seen from the $L=\infty$ curve for $N_c=1$ in Figure \ref{fig:energy}.

\subsection{Outline}

Using the general orientation to the three $J$-regions provided by these observations on the energy
gain $\Delta{E}(J)$, in Section \ref{sec_overall} we will give an overview of the types of
situations encountered in this study (fixed number, fixed small density), and explain how these
connect to the usual thermodynamic limit and well-known finite density results.  In Section
\ref{sec_RG} we present renormalization group arguments for the $\Delta{E}(J)$ behavior at finite
but small densities.  Section \ref{sec_ultralow} considers the ultralow density situation of fixed
$N_c$.  In Section \ref{sec_1el}, we detail the case of a single electron ($N_c=1$), which is exactly
solvable.  Results for finite numbers of electrons, $N_c>1$, are outlined in Section \ref{sec_few}.  Since
our numerical calculations (exact diagonalization) are restricted to smaller $L$ at larger $N_c$,
the results for $N_c=3,5,\ldots$ also provide a description of how the half-filling case is
approached, e.g., how spin and charge are more and more decoupled at larger fillings.  In Section
\ref{sec_2D_3D}, we consider briefly the case of higher dimensions.


\section{\label{sec_overall} Overview; fixed number versus fixed small densities}

\begin{figure}[tb]
\begin{center}
  \includegraphics[width=1.05\columnwidth]{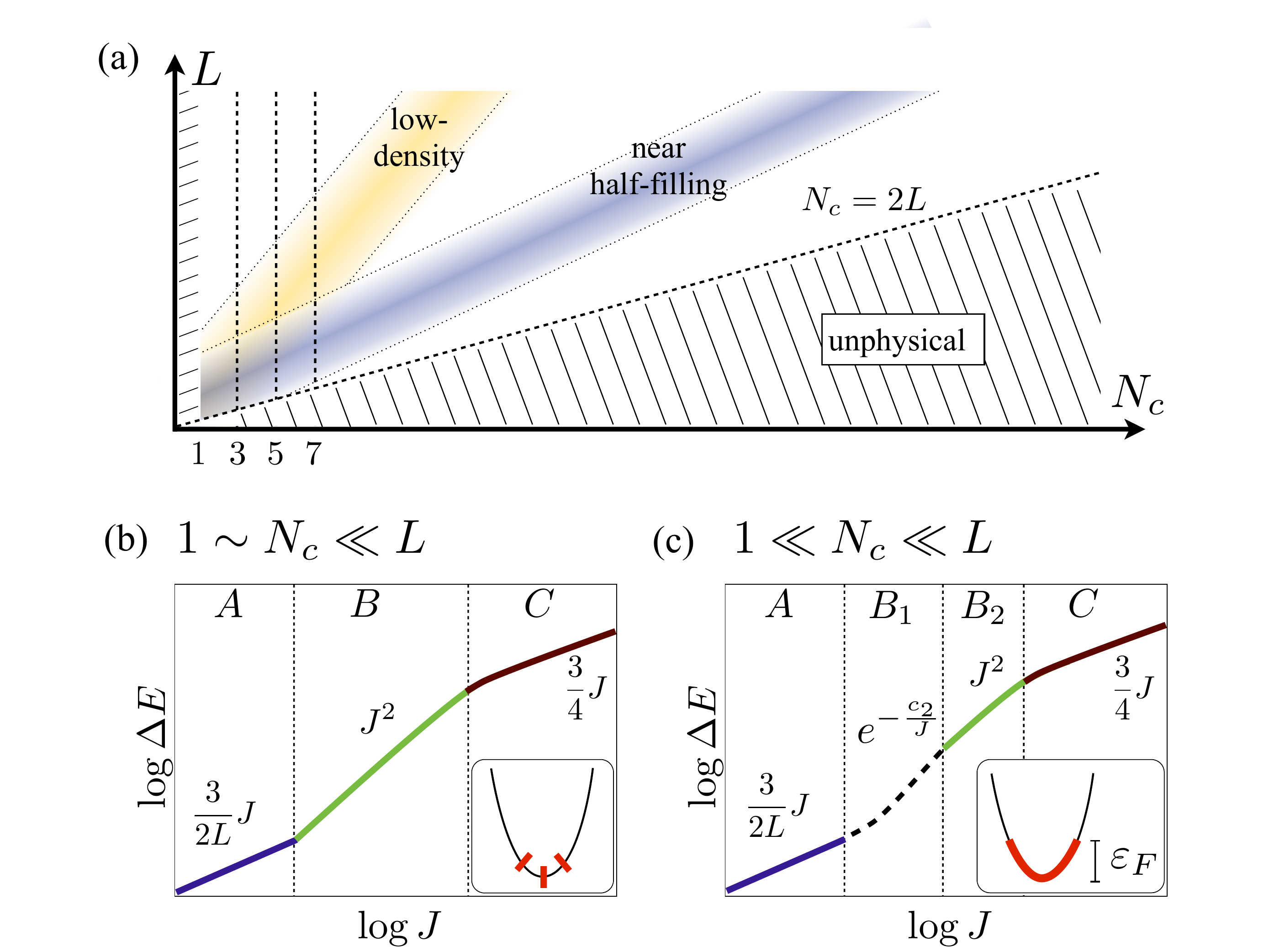}
\end{center}
  \caption{\label{fig:regimes_cartoon} 
(a) Different regimes on the $N_c$-$L$ parameter space: the
    ultralow-density regime ($N_c$ fixed, vertical lines), the low-density regime ($N_c/L$ fixed at
    small value, steep shaded area) and regime near half filling ($N_c/L\sim1$, shaded area with
    smaller slope).
(b,c) schematics showing different $\Delta{E}(J)$ behaviors in different $J$ regions, for the cases
    of small and large $N_c$, respectively.  See text for details.
}
\end{figure}

Low-density and ultralow-density situations correspond to different orders of limits for system size
$L$ and particle number $N_c$.  Figure \ref{fig:regimes_cartoon}(a) charts out several different
regimes.  The usual thermodynamic limit involves both $L\to\infty$ and $N_c\to\infty$ while keeping
the density $(N_c/L)$ fixed.  On the $N_c$-$L$ plane, this corresponds to going toward large sizes
along a finite-slope path, e.g., along one of the shaded paths.  Low but finite density corresponds
to a steep but finite-slope path.  The ultralow-density situation involves fixed $N_c$ at arbitrary
$L$, including $L\to\infty$, these are vertical lines in Figure \ref{fig:regimes_cartoon}(a) and do not
correspond to the regular thermodynamic limit.

We note that the regimes overlap at smaller $N_c$, e.g., the vertical dashed lines cross through the
shaded areas indicating the thermodynamic limit.  Thus, for finite but large $L$ and finite
$N_c\ll{L}$, it may be ambiguous to decide whether the physics of the system is best described by
low-density results in the thermodynamic limit, or by the ultralow-density physics that does not
correspond to the usual thermodynamic limit.

In  Figure \ref{fig:regimes_cartoon}(b,c) we have summarized our results for the energy gain (analogue of
the Kondo temperature) in the 1D case; the derivations will appear in later sections. Differences
occurring for higher-dimensional baths will be discussed in Sec.~\ref{sec_2D_3D}.

Panels \ref{fig:regimes_cartoon}(b) and \ref{fig:regimes_cartoon}(c) correspond to $1\sim N_c \ll
L$ and $1\ll N_c\ll L$, respectively. The insets show a rough ``band filling'' picture of these cases:
for small $N_c$ one may think of a finite number of single-particle levels of the conduction band being
occupied, while for larger $N_c$ a picture of small band filling is more appropriate.

In both cases and for finite $L$, there is a finite-size-dominated A region at small $J$ where
$\Delta{E}=c_1J$, with $c_1=\frac{3}{2L}$ in every case except for $N_c=1$, where we have
$c_1=\frac{3}{4L}$.
The C region with localized singlet and energy gain $\Delta{E}=\frac{3}{4}J$ is
also the same for all cases. These behaviors in A and C regions hold not only for the cases we are
considering but even for the half-filling situation.

For the interesting B region, $\Delta{E}(J)$ is nonperturbative.  For fixed small $N_c$, the behavior
is $\Delta{E}\sim{J}^2$. For $N_c=1$, the exact solution yields
$\Delta{E}=\left(\frac{3}{8}J\right)^2$. There are no exact solutions for fixed $N_c>1$, but we present
strong numerical evidence that the energy gain in the B region is identical for $N_c=3$, and we
conjecture that this applies for all small odd $N_c$.

For larger $N_c$, shown in Fig. \ref{fig:regimes_cartoon}(c), the B region can have two types of
behavior.  If $J$ is small enough that the Kondo energy scale is significantly smaller than the Fermi
energy with respect to the band bottom, screening is dominated by electron states around the Fermi
points where the dispersion is nearly linear, so that we recover the usual Kondo effect, and one
expects a $J$ dependence according to $\Delta{E}(J) \sim e^{-c_2/J}$.  On the other hand, if $J$ is
larger so that the Kondo energy scale is of the order or larger than the Fermi energy, linearization is
not possible and the density of states at the bottom of the spectrum has to be taken into account.  For
the 1D conduction bath, this turns out to lead to $\Delta{E}\sim{J}^2$, as described in Section
\ref{sec_RG}. We have marked these two regions as B$_1$ and B$_2$ in Figure
\ref{fig:regimes_cartoon}(c).

The limit $L\to\infty$ with $N_c$ fixed yields the ultralow-density limit announced in the
introduction; in this limit only the B (or B$_2$) and C regions survive. In the low-density limit, with
$L\to\infty$ and $N_c\to\infty$, there is a B$_1$ region at small $J$ which connects to the standard
(nonperturbative) Kondo setting with $\Delta{E}(J) \sim e^{-c_2/J}$. In contrast, the B$_2$ region
occurring for larger $J$ does not have the familiar $e^{-c_2/J}$ behavior for the energy gain (in 1D),
but nevertheless has nonperturbative behavior.


\section{\label{sec_RG} Renormalization Group predictions for small finite densities}

Results for the low-density case in the thermodynamic limit can be obtained using a perturbative
renormalization-group (RG) expansion around the free-moment fixed point of the Kondo model, i.e., a
generalization\cite{withoff,fv04} of Anderson's poor man's scaling.\cite{poor} We note that the results
continue to apply for finite $L$ as long as the bath level spacing is small compared to the Kondo
temperature $\TK$.

We restrict our attention to the case of a one-dimensional conduction band as in
Eq.~\eqref{eq:minimalKondoHam}. It is convenient to work in a grand-canonical ensemble,
with a chemical potential $\mu$; in the following $\mu$ and all energies will be measured
relative to the band bottom. The density of states at small energies is
\begin{equation}
\label{dos1}
\rho(\omega) = \rho_0 \left|\frac{\omega}{D}\right|^{r} \Theta(\omega),~~ r=-1/2
\end{equation}
with $D=4$ the bandwidth and $\rho_0=1/(4\pi)$, such that chemical potential and average filling
$n_c=N_c/L$ are related by
\begin{equation}
\label{ncmu}
n_c \propto \mu^{1/2}.
\end{equation}

Eqs.~\eqref{eq:minimalKondoHam} and \eqref{dos1} define an unconventional Kondo problem,
with non-constant and strongly asymmetric density of states. In the course of the RG, a
scattering potential $V$ at the impurity site will be generated which keeps track of the
particle--hole asymmetry. For the limiting case of $\mu=0$ the weak-coupling beta
functions for the dimensionless running couplings $j=\rho_0 J$ and $v=\rho_0 V$
read\cite{zawa_asy,florens07}
\begin{align}
\frac{d j}{d \ln \Lambda} = r j - \frac{j^2}{2} + 2 v j\,,~~~
\frac{d v}{d \ln \Lambda} = r v + \frac{3j^2}{16} + v^2
\label{poor}
\end{align}
to second order, with $\Lambda$ being the running UV cutoff, and the initial values
$j_0 = \rho_0 J$ and $v_0=0$ according to Eq.~\eqref{eq:minimalKondoHam}.

\subsection{Kondo scale}

Clearly, for small $j$ and $v$, the first (tree-level) term in the beta function dominates the flow.
For $r=-1/2<0$ the initial $j$ grows under RG and diverges at $\Lambda=\TK$ with\cite{mitchell13}
\begin{equation}
\label{tkb2}
\TK
\propto D \left(\frac{\rho_0 J}{-r}\right)^{-1/r} = D \left(2{\rho_0}J\right)^{2}
\end{equation}
As usual, the energy scale $\Lambda=\TK$ where $j$ diverges is the estimate for the Kondo energy or
temperature scale, expected to be proportional to the energy gain $\Delta{E}$ introduced earlier.
This $\mu=0$ result continues to hold for nonzero $\mu$ as long as $\TK>\mu$, as in this case the RG
flow reaches strong coupling before the deviation from the band edge becomes relevant -- this is
exactly what defines the region B$_2$ in Fig.~\ref{fig:regimes_cartoon}.

In the opposite limit $\TK\ll\mu$, corresponding to region B$_1$, the dominant contribution to
screening arises from the regime $\Lambda<\mu$ where the density of states can be approximated as
constant, $\rho(\mu) = \rho_0 (\mu/D)^{-1/2} \propto \rho_0/n_c$.

The standard exponential estimate for the energy scale applies, consequently:
\begin{equation}
\label{tkb1}
\TK \propto D \exp\left(-\frac{c_2}{J}\right)
\end{equation}
with $c_2 \propto n_c/\rho_0$.

\subsection{Crossovers between the A, B, and C regions}

These considerations allow to extract the locations of the crossovers between the various regions.
First, the B--C crossover is set by $\TK\sim 1$, which implies $\rho_0 J_{BC} \sim 1$.

Second, the boundary between the B$_1$ and B$_2$ regions is defined by $\TK \sim \mu$; using
Eqs.~\eqref{ncmu} and \eqref{tkb2} this yields $\rho_0 J_{B12} \sim n_c$. Alternatively, we can demand
the $\TK$ expressions in Eqs.~\eqref{tkb2} and \eqref{tkb1} to match, resulting in $\rho_0 J_{B12}
\ln[1/(\rho_0 J_{B12})] \sim n_c$ which is identical to the first criterion up to logarithmic accuracy.
This shows how the region B$_1$ disappears when passing from the low-density to the ultralow-density
limit, $n_c\to 0$.

Third, the boundary of the A region can also be found be equating the expressions for $\TK$ (or $\Delta
E$). Using $\Delta E \propto J/L$ in the A region, this yields for a direct crossover from A to B$_2$
the relation $\rho_0 J_{AB2} \sim 1/L$. In contrast, the A--B$_1$ crossover occurs at $\rho_0 J_{AB1}
\sim N_c/ (L \ln L)$; this formula is valid only for $N_c\gg 1$ and up to logarithmic accuracy.

Taken together, these relations show that $1\sim N_c \ll L$ (with $L$ finite) implies a direct
crossover from A to B$_2$ as in Fig.~\ref{fig:regimes_cartoon}(b), whereas for large $N_c$ a B$_1$
region intervenes.


\section{\label{sec_ultralow} Fixed number of conduction electrons (ultralow densities)}

In this section we describe systems with a fixed number $N_c$ of electrons in the one-dimensional
conduction bath.  We will restrict to odd numbers of $N_c$ so that the total spin is integer and
hence can be a singlet.
In Section \ref{sec_1el} we treat analytically the $N_c=1$ case and describe the energy gain and
spatial profiles.  In Section \ref{sec_few} we describe $N_c>1$, again focusing on the energy gain and
spatial profiles.  Section \ref{sec_pert} describes perturbative results for the A (small $J$) region
for finite $L$.

\subsection{\label{sec_1el} Exactly solvable case, $N_c=1$}

Focusing on the solvable case of a single fermion in the bath ($N_c=1$), we will derive below the
energy gain in B and C regions.  We also show that the impurity-bath spin-spin correlator
$\chi_j=\xpct{ \vec{S}_{\mathrm{imp}}\cdot \vec{s_j}}$ is locked to the density profile
$n_j=\sum_{\sigma}\xpct{c_{j\sigma}^{\dagger}c_{j\sigma}}$, through the relation
$\chi_j=-\frac{3}{4}n_j$.

The three $J$-regions have simple spatial interpretations in terms of the density profile of the single
electron.  In an infinite chain, in the ground state, the fermion is localized around the
impurity-coupled site ($i=0$) with localization length $\xi$.  ($\xi$ decreases with increasing
$J$.)  At large $J$ (region C), the itinerant fermion is almost completely localized at site 0
($\xi\lesssim1$).  At smaller $J$, the itinerant fermion is spread over multiple sites $\xi>1$
(region B).  In an infinite system, this region would extend to arbitrarily small $J$.  However, for
any finite size $L$, there is a boundary-sensitive small-$J$ region (region A) where the fermion
cloud extends over the whole system ($\xi\gtrsim{L}$).

\subsubsection{Analytic solution for energy gain}

To examine the ground state, we restrict to the singlet sector.  Within this sector, the states can
be written in the basis of single-particle momentum eigenstates
\begin{equation}
\ket{k}= \frac{1}{L\sqrt{2}} \sum_j e^{-ikj}  \Big( \ket{\uparrow;j\downarrow}-
\ket{\downarrow;j\uparrow} \Big) \, .
\end{equation}
The $\ket{S;js}$ notation refer to the impurity spin $S$ and the electron position $j$ and spin $s$.
Within the singlet sector the Kondo coupling serves as a local potential on the site $j=0$, of
strength $-\frac{3}{4}J$, i.e., this sector is described by a single-particle problem in a
resonant level model:
\begin{equation}
H ~=~ -2\sum_{k}\ket k\cos k \bra k-\frac{3}{4}\frac{J}{L}\sum_{k,k'}\ket k\bra{k'}
\end{equation}
The components $\psi_{k}$ of an eigenstate $\ket{\psi}=\sum_{k}\psi_{k}\ket k$
with energy $E$ obey the relation
\begin{eqnarray}
\left(-2\cos k -E\right)\psi_{k} & = & \frac{3}{4}\frac{J}{L}\sum_{k'}\psi_{k'}
\end{eqnarray}
Summing over $k$ one gets closed equations for $E$ and for $\psi_k$
\begin{eqnarray}\label{eq:A2EigenEnergy}
1 & = & \frac{3}{4}\frac{J}{L}\sum_{k}\left(-2\cos k-E\right)^{-1} \\
\label{eq:1elcoeffs}
\psi_{k} & = & \frac{1}{\mathcal{N}}\left(-2\cos k-E\right)^{-1}
\end{eqnarray}
where $\mathcal{N}$ is a normalization constant.  This gives all the singlet eigenstates at finite $L$.
For $L\to\infty$ the continuum version of Eq.\ \eqref{eq:A2EigenEnergy} gives the ground-state energy,
because the rest of the states are part of a continuum in $E\in\left[-2,2\right]$.  The ground state
energy $E_0$ thus satisfies $\frac{4}{3J} = -\int\frac{dk/(2\pi)}{2\cos k+E_0}$, leading to $E_0=
-\sqrt{\left(\tfrac{3}{4}J\right)^{2} + 4}$, so that the energy gain, $\Delta E = -2-E_0$, is
\begin{equation}
\Delta{E}  = -2 ~+~ \sqrt{\left(\tfrac{3}{4}J\right)^{2} + 4 }\label{eq:A2EnergyVsJk} ~\approx~
\begin{cases} \frac{9}{64}J^2 & J\ll1  \\  \frac{3}{4}J & J\gg1 \end{cases}
\end{equation}
This is the solid curve in Figure \ref{fig:energy}.  For finite $L$, the $\frac{9}{64}J^2$ behavior
gets cut off at smaller $J$ and is replaced by the A region, $\Delta{E}\approx\frac{3}{2}J$.  This can
be numerically extracted from Eq.\ \eqref{eq:A2EigenEnergy} or can be calculated perturbatively
(subsection \ref{sec_pert}).

\subsubsection{Real-space profiles; density and spin correlator}

\begin{figure}[tb]
  \includegraphics[width=0.99\columnwidth]{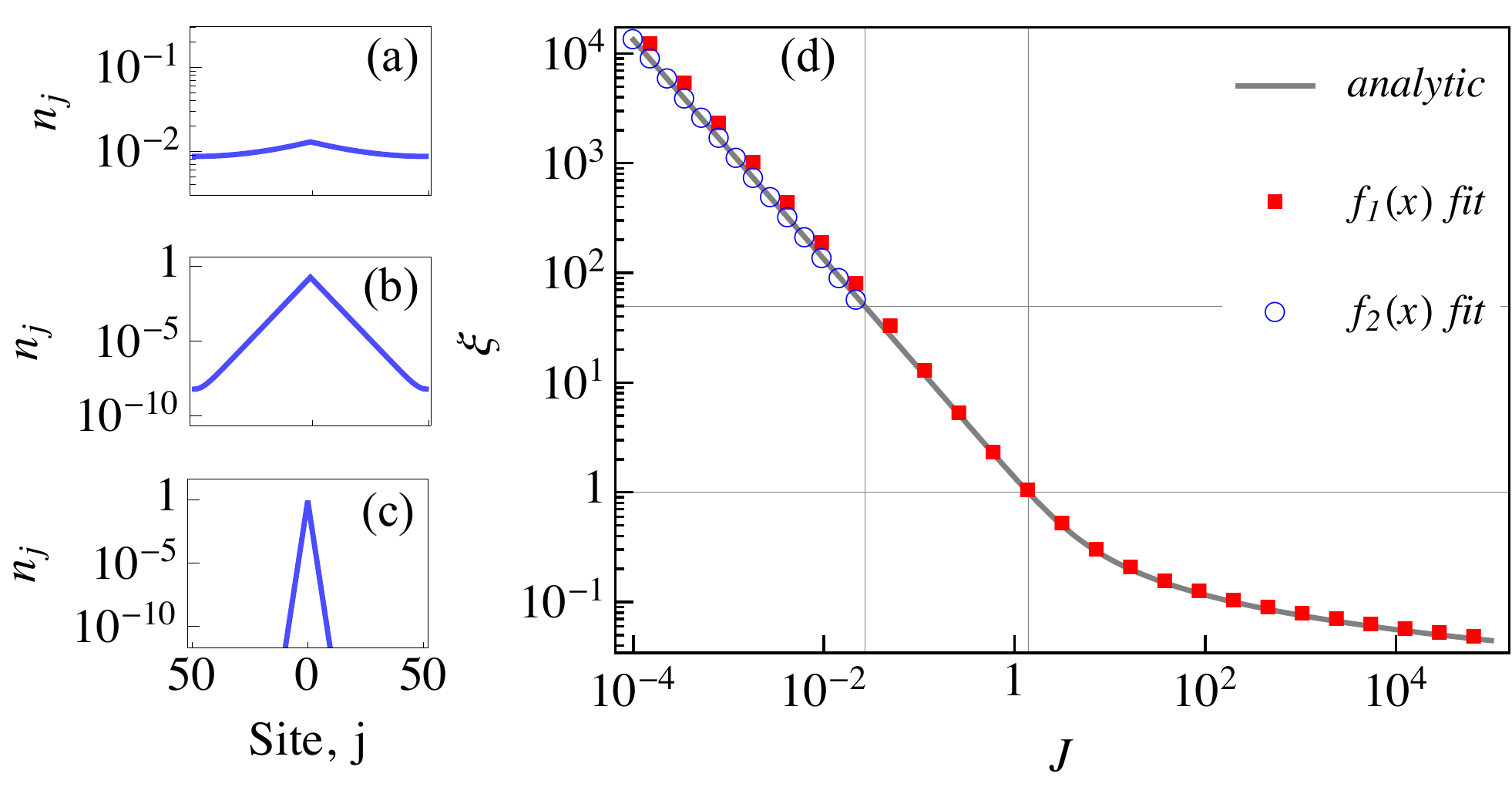}
  \caption{\label{fig:1el_exp}
Ground state for single bath electron ($N_c=1$), $L=100$.  (a-c) Density profiles in A, B, C regions: $J=0.02$;
$J=0.5$; $J=5$.
(d) The localization length  $\xi$.  Black solid line is analytic large-$L$ expression
\eqref{eq:1el_exp}.  Filled dots are obtained by fitting function $f_1(x)$ (see text) to
$(\frac{L}{2}-1)$ sites around the impurity.  Open dots in the A region are obtained by fitting
$f_2(x)$.
}
\end{figure}

The density is
\begin{eqnarray}\label{eq:1el_density}
n_j & = & \sum_{s}\av{c_{j,s}^{\dagger}c_{j,s}}=\abs{\psi_{j}}^{2}= \abs{\frac{1}{\sqrt{L}}\sum_{k}e^{-ikj}\psi_k}^{2}
\end{eqnarray}
with $\psi_k$ given by Eq.\ \eqref{eq:1elcoeffs}.  The impurity-bath spin-spin correlator is
\begin{eqnarray} \label{eq:1el_spin}
\chi_j & = & \av{\vec S_\mathrm{imp} \cdot \vec s_{j}}=-\frac{3}{4}\abs{\psi_{j}}^{2}=-\frac{3}{4}n_{j} \, ,
\end{eqnarray}
showing that the two quantities are locked to each other in the $N_c=1$ case.
In the $L\to\infty$ limit, one obtains $n_j\propto e^{-|j|/\xi}$, with
\begin{equation}\label{eq:1el_exp}
\xi^{-1}  = -\ln\left[\sqrt{\left(\tfrac{3}{8}J\right)^{2}+1}- \tfrac{3}{8}J \right].
\end{equation}
For finite $L$, the density (and $\chi_j$) profile is exponentially localized around the impurity
site in the B and C regions, but is modified by the boundary in the A region, as shown in Figure
\ref{fig:1el_exp}(a-c).  Figure \ref{fig:1el_exp}(d) compares the expression \eqref{eq:1el_exp} with
the length scale $\tilde{\xi}$ obtained from an exponential fit $f_1(x)=A_1\exp[-x/\tilde{\xi}]$ in
the B and C regions ($J\gtrsim1$) for a $L=100$ system.  In the A region, we find that the
real-space profile is well-described by the corrected form $f_2(x)= A_2\exp[-x/\tilde{\xi}+x^2/(\tilde{\xi}L)]$; Figure \ref{fig:1el_exp}(d) also compares $\tilde{\xi}$ obtained with this fit in the A
region.

\subsubsection{Entanglement entropy profile}

We will now characterize the spatial structure of the system using block-block entanglement entropy,
$S(J,r)=-\text{Tr}\left[\rho_{r}\ln\rho_{r}\right]$, where $\rho_r$ is the reduced density matrix of a
block containing the impurity spin and the $(2r-1)$ sites centered around the impurity-coupled site.
This is analogous to studies in finite-density impurity systems where block entanglement entropies have
been used to describe the real-space impurity screening cloud. \cite{SorensenLaflorencieAffleck,
SaleurVasseur_PRB13, GhoshRibeiroHaque_2013}

\begin{figure}[tb]
\includegraphics[width=1\columnwidth]{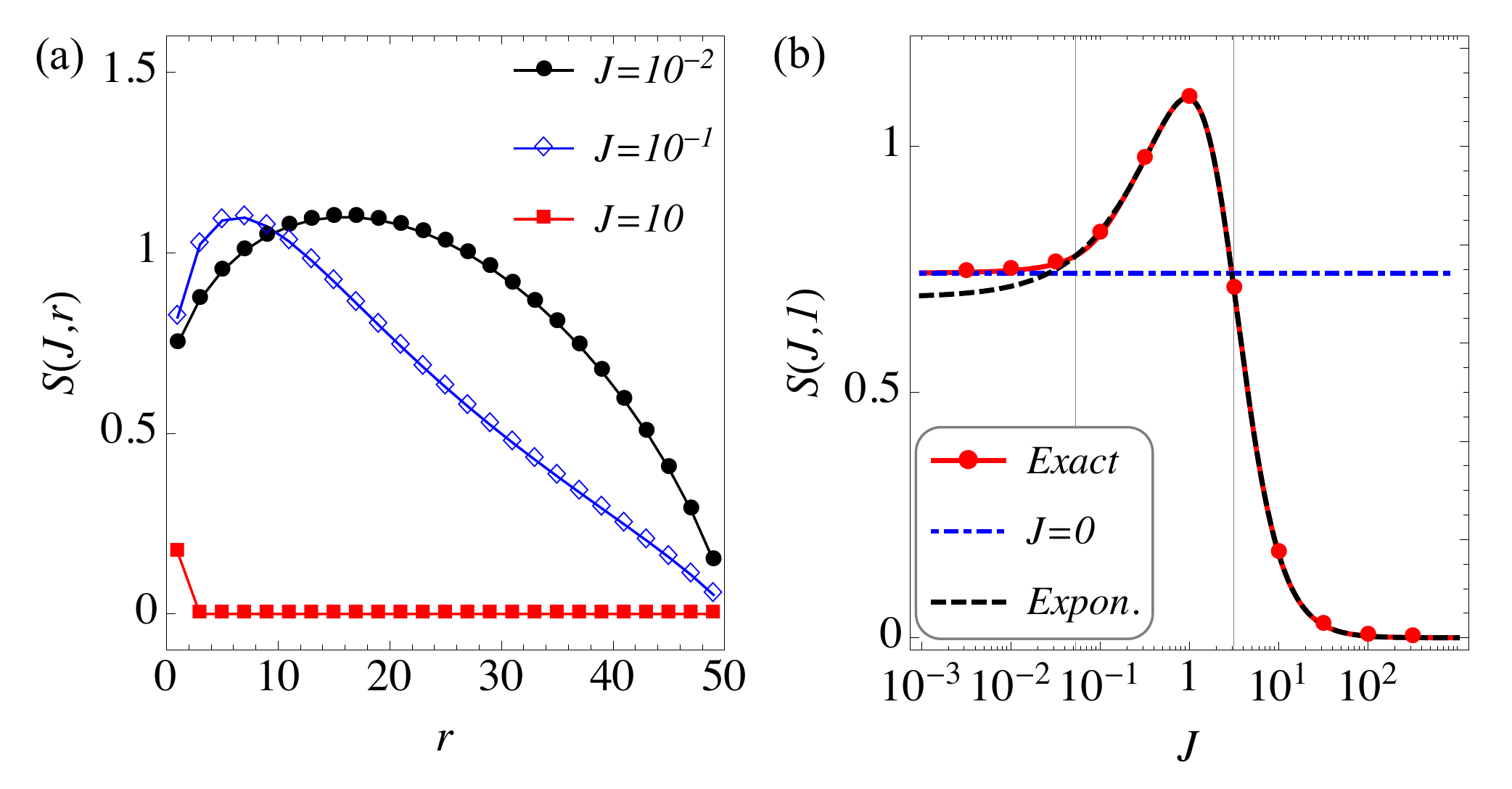}
  \caption{\label{fig:entangl_ent}
Entanglement entropy $S(J,r)$ for $N_c=1$, $L=100$, between a block containing the impurity and the
$(2r+1)$ bath sites around it, and the rest of the system.
(a) $S(J,r)$ against block size $r$, for one $J$ value each in the A, B, C regions.
(b) $S(J,1)$ against $J$.  Exact numerical curve crosses over from the constant $S(0,1)$ value to
the curve $S_{\mathrm{Expon}}(J,1)$ calculated assuming exponentially localized electron,
Eq.\ \eqref{eq:BBE_uni} to Eq.\ \eqref{eq:BBE_exp}.  Vertical lines demarcate A, B, C regions.
}
  \end{figure}

Figure \ref{fig:entangl_ent}(a) shows the typical behaviors of the entanglement entropy as a
function of block size, for $J$ values in the A, B, C regions.  In the C region, $S(J,r)$ is nearly
zero for all $r>1$, since the electron is localized at the impurity-coupled site.  In the B region,
there is structure indicative of the localization length $\xi(J)$.  The A region curve is the
entanglement entropy of a uniform system; the lack of left-right symmetry is due to the block
containing the  impurity spin  and therefore being inequivalent to its complement when
containing half the lattice sites.

In panel (b) the entanglement entropy is shown for block size $r=1$, i.e., the block contains only
the impurity and the site $j=0$.  In the extreme C region the cloud is completely localized on this
site, so that the rest of the system is decoupled, $S_{J\to\infty}(J,r)\to0$. In the extreme A
region the density is uniform, so that the entanglement entropy has constant $L$-dependent value.
Constructing the density matrix explicitly, we obtain for this uniform case
\begin{equation}\label{eq:BBE_uni}
S(J=0,r=1)= \frac{L-1}{L} \log\left(\frac{2L}{L-1}\right) + \frac{1}{L}\log\left(L\right).
\end{equation}
%
%
%
In the B and C regions, the electronic cloud decays exponentially $n_j\propto e^{-j/\xi(J)}$ where
$\xi(J)$ is given in Eq.~\eqref{eq:1el_exp}.  Explicit calculation gives
%
%
\begin{equation} \label{eq:BBE_exp}
S_{\mathrm{Expon}}(J,r=1) ~=~ - 2M_1 \log{M_1} ~-~  M_2 \log{M_2} \,  ,
\end{equation}
with $M_1(J)=\left[1+e^{1/\xi(J)}\right]^{-1}$ and $M_2(J)=\tanh\left(\frac{1}{2\xi(J)}\right)$.
Figure \ref{fig:entangl_ent}(b) shows the above expressions together with the exact numerical
$S(J,1)$ for $L=100$.  The exact curve moves from $S(0,1)$ to $S_\mathrm{Expon}(J,1)$ as $J$ is
increased from the A to the B region; the two curves cross near $J=\frac{16}{3L}$, which is the
boundary between A and B regions at large $L$ as obtained from the condition $\xi(J)=L/2$.

\subsection{\label{sec_few} Few mobile electrons; $N_c>1$}

We now look at the energy gain and the real-space profiles for a few fermions ($N_c$ odd and $>1$) in a
1D bath. As we shall show, a general feature is that, for  $L\to\infty$, the $N_c>1$ systems behave in
some ways similar to the $N_c=1$ system.  We will see this both in the energy gain and in the
real-space profiles. Intuitively, the reason is that for large $L$ the ground state involves one of the
bath fermions localized around the impurity while the other fermions spread out with vanishing density
and therefore negligible effect.  Thus the physics of the impurity interacting with a single fermion is
dominant, so that one recovers signatures of the energy gain $\Delta{E}$ and localization length $\xi$
derived in Section \ref{sec_1el}.

\subsubsection{Energy gain}

Unlike the $N_c=1$ case, it is not possible to obtain an analytical expression or simple equation for
$\Delta{E}$.  However, numerical calculation (Figure \ref{fig:energy}) shows that there are three $J$
regions with the same characteristics.  For large $J$ (C region) the energy gain is the singlet energy
$\Delta{E}\approx\frac{3}{4}J$.  At small $J$ (A region) the energy gain is perturbative and
$L$-dependent, $\Delta{E}\sim\frac{3}{4L}J$ for any $N_c>1$ (Section \ref{sec_pert}).  This is half the
energy gain for $N_c=1$ in the A region.

The intermediate $J$ region (B region) is more tricky to characterize; it is not straightforward to
infer the $J$-dependence in the B region by looking only at the numerical $\Delta{E}(J)$ with
available sizes.  Analyzing instead the second derivative $\partial_{JJ}\Delta E(J)$, we provide
numerical evidence that in the $L\to\infty$ limit the $\Delta{E}(J)$ curve for $N_c>1$ coincides
with the $\Delta{E}(J)$ curve calculated for $N_c=1$.  This implies that the B region is described by
$\Delta{E}\sim\frac{9}{64}J^2$ also for $N_c>1$.

\begin{figure}[tb]
\begin{center}
  \includegraphics[width=1\columnwidth]{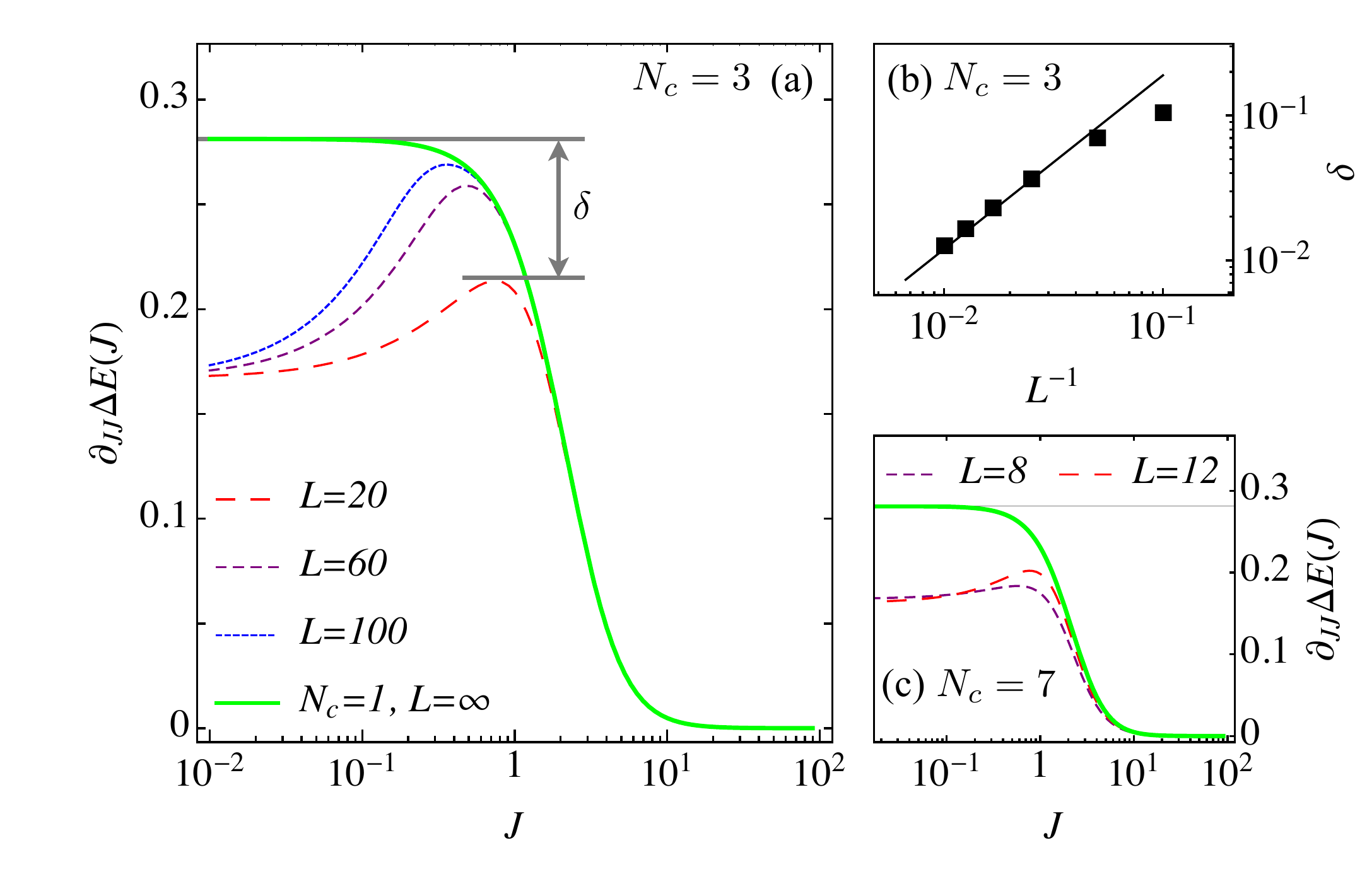}
\end{center}
  \caption{\label{fig:energy_gain_2d_derivative}
(a) Second derivative of the energy gain $\partial_{JJ}\Delta E(J)$ for $N_c=3$ mobile electrons and
    different $L$.  Thick solid line shows the exact $\partial_{JJ}\Delta E(J)$ for $N_c=1$,
    $L\to\infty$, Eq.~\eqref{eq:A2EnergyVsJk}.  With increasing $L$, the 3-electron curves clearly
    approach the exact 1-electron curve.
(b) Difference between maximum of the  $N_c=3$ curves $\partial_{JJ}\Delta E(J)$ and the maximum of
    the  $N_c=1$ curve, against $L$.
(c) Similar data as (a), but for $N_c=7$ electrons.
}
  \end{figure}

Figure \ref{fig:energy_gain_2d_derivative}(a) shows the second derivative for $N_c=3$, for system sizes
from $L=10$ to $L=100$.  Clearly, $\partial_{JJ}\Delta E(J)$ approaches the $L\to\infty$ solution of
the single-electron ($N_c=1$) case.  In Figure \ref{fig:energy_gain_2d_derivative}(b) we show the
difference between the maximum value of $\partial_{JJ}\Delta E(J)$ of the finite-size $N_c=3$ data,
from the  exact $N_c=1$ solution.  The difference decreases with $L$, apparently with a super-linear
power law.  This provides relatively strong evidence that, in the limit $L\gg \infty$, the B region for
$N_c=3$ has identical $\Delta{E}(J)$ behavior as for the exactly solved $N_c=1$ case.

For larger $N_c$, it is difficult to use large enough $L$ for a proper finite-size scaling analysis.
However, Figure \ref{fig:energy_gain_2d_derivative}(c) shows the second derivative $\partial_{JJ}\Delta
E(J)$ for $N_c=7$ conduction electrons in a $L=8$ bath and in a $L=12$ bath. The features of these
curves, and the way in which they approach the analytic $N_c=1$ curve with increasing $L$, are very
similar to the $N_c=3$ case.  This leads us to suggest that for \emph{any} fixed odd $N_c\geq3$, the
B$_2$ region is described by $\Delta{E}\sim\frac{9}{64}J^2$. As discussed in Section~\ref{sec_RG},
for $N_c\gg 1$ there will be an additional B$_1$ region at small $J$ with exponential behavior of
$\Delta E$; this is not captured by our finite-size numerics.

\begin{figure}[tb]
  \includegraphics[width=1\columnwidth]{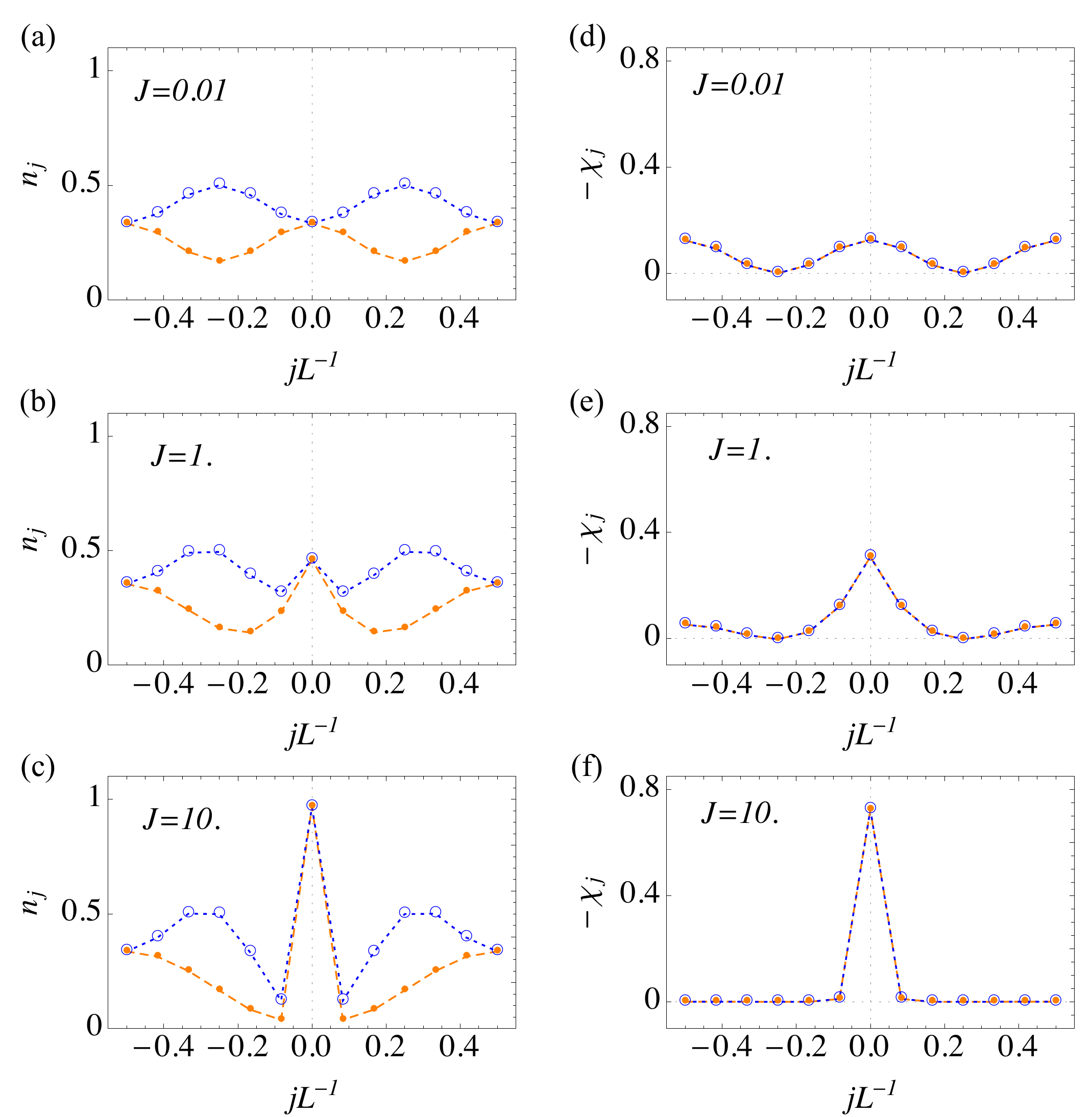}
  \caption{\label{fig:real_space} Real space profiles in an $L=12$ bath: (a-c)   electronic density
    $n_j$;   (d-f) impurity-bulk spin susceptibility $\chi_j$.
Each panel shows data for   $N_c=3$  (filled orange circles) and $N_c=5$ (empty blue circles).
  }
  \end{figure}

\subsubsection{Real space profiles; density and spin correlator}

Figure \ref{fig:real_space} plots the spatial dependence of the total electronic density $n_j$ and the
impurity-bulk spin correlator $\chi_j$ for $N_c=3$ and $N_c=5$, with $L=12$ sites.

For $J\ll{1}$, the densities are primarily determined by the bath.  The single-particle spectrum of the
bath includes two degenerate states at momentum $k=0$, and four degenerate states each at momenta
$k=\pm\frac{2\pi}{L}n$. Thus the cases of $N_c=4n-1$ and $N_c=4n+1$ are closely linked, corresponding
to having one or three states filled among the highest 4-fold degenerate set. This explains why the
density profiles for $N_c=3$ and $N_c=5$ are closely linked.   The density profiles for $N_c=7$ and
$N_c=9$ form a similar pair. More strikingly, the correlator $\chi_j$ are identical for $N_c=4n-1$ and
$N_c=4n+1$ in the $J\ll{1}$ limit.  These features in the perturbative $J\ll1$ region will be explained
in more detail in Section \ref{sec_pert}.

Similar arguments can be made in the large-$J$ limit, where one particle is bound to the impurity and
the remaining $(N_c-1)$ fermions can be treated as free fermions in an open-boundary $(L-1)$-site
chain. Again, the $N_c=3$ and $N_c=5$ density profiles are closely linked, and the spin correlators are
nearly identical.  Surprisingly, we find $\chi_j$ for $N_c=4n-1$ and $N_c=4n+1$ to be nearly identical
even for intermediate $J$ (Figure \ref{fig:real_space} middle row), where we cannot use free-fermion
ideas to explain this feature.

For large $L$, the real-space behaviors for finite $N_c>1$ is governed by the single-electron
localization length.  In Figure \ref{fig:real_space} we illustrate this through the spin correlator
$\chi_j$.  We extract the length scale of localization around the impurity by fitting $\abs{\chi_j}$
to $f_1(x)=A_1\exp[-x/\xi_{\chi}]$ in the B and C regions, using $\chi_j$ on only three sites near
the impurity to avoid complications such as sign changes of $\chi_j$ at larger $j$.  For small $J$,
it is necessary to incorporate the boundary with the modified exponential $f_2(x)=
A_2\exp[-x/\tilde{\xi}+x^2/(\tilde{\xi}L)]$, as in the 1-electron case.  In addition, from Figure
\ref{fig:real_space} (top row) it is clear that an overall $2k_F$ oscillation is important at small
$J$.  Therefore for small $J$ a fit with $f_2(x)\cos^2(2\pi{x}/L)$ is used to extract the length
scale $\xi_{\chi}$.  In this case about 35\% of the sites are used for the fit.

\begin{figure}[tb]
\centering  \includegraphics[width=0.98\columnwidth]{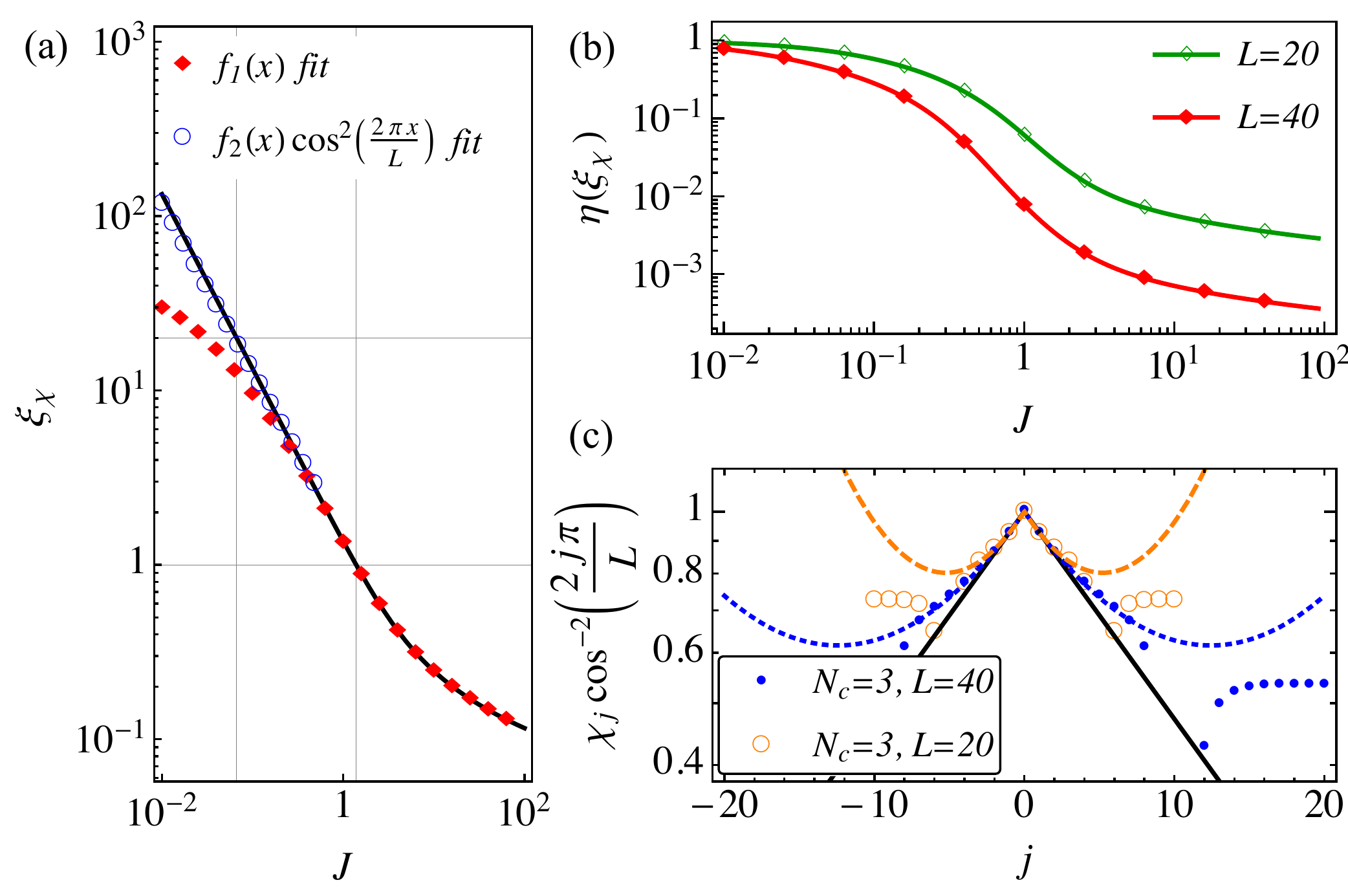}
  \caption{\label{fig:3el_exponent}
(a) Decay length scale of $\chi_j$ close to the impurity, for $L=40$,
    $N_c=3$.  Length scale $\xi_\chi$ obtained from  $f_1(x)$ fit (red dots), and from
    $f_2(x)\cos^2(2\pi{x}/L)$ fit (open blue squares) are shown.  Localization length for $N_c=1$
is given by solid line.
(b) The relative difference between $N_c=3$ and $N_c=1$ electron exponents $\eta(\xi_\chi)$ is shown
    for $L=40$ and $L=20$.  Here $\xi_\chi$ is obtained by fitting with $f_1$.
(c) For $J=0.1$, the exact numerical $\chi_j$ data are shown together with best fits with
    $f_2(x)\cos^2(2\pi{x}/L)$.  The oscillatory factor is divided out.  Exact data are shown as open
    orange circles ($L=20$) and filled blue circles ($L=40$); fits are shown with orange dashed line
    and blue dotted lines.  Black solid line shows $\chi_j$ for  $N_c=1$, $L=40$.
}
\end{figure}

Figure \ref{fig:3el_exponent}(a) compares the length scale $\xi_{\chi}$ extracted numerically from the
3-electron $\chi_j$, with the 1-electron result $\xi^\mathrm{1el}$ from Eq.\ \eqref{eq:1el_exp}. The
close match indicates that the localization is governed by the $N_c=1$ length scale.  Figure
\ref{fig:3el_exponent}(b) shows that this match gets better with increasing $L$, by plotting the
relative difference, $\eta(\xi_\chi) = \left|\xi_{\chi}-\xi^\mathrm{1el}\right|/\xi_\mathrm{1el}$,
between the 1-electron length scale and the 3-electron length scale obtained from a $f_1(x)$ fit. (The
$f_1$ fit is not expected to be reasonable at small $J$.)
Figure \ref{fig:3el_exponent}(c) shows that $f_2(x)\cos^2(2\pi{x}/L)$ is a meaningful description for
moderate distances from the impurity.

\begin{figure}[b]
  \includegraphics[width=0.98\columnwidth]{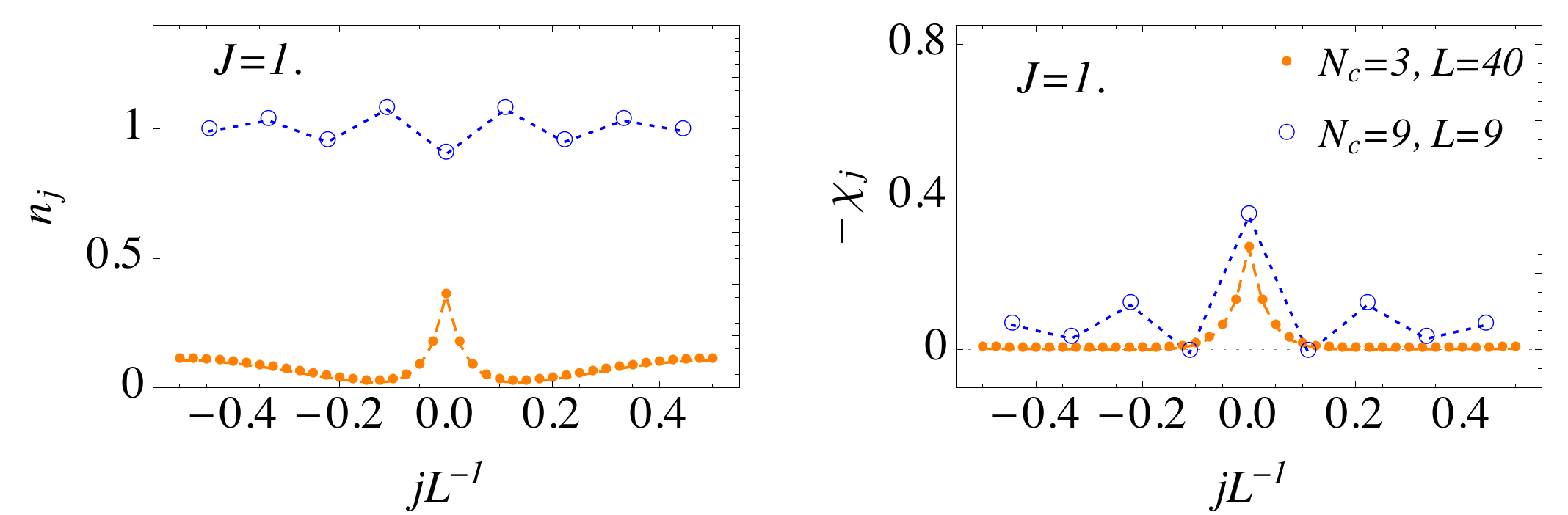}
\caption{\label{fig:3el_9el}
Spin-charge decoupling upon approaching half filling.  A $N_c{\ll}L$ case and a $N_c=L$ case are
shown.
}
\end{figure}

Figure \ref{fig:3el_9el} shows how the density $n_j$ and the spin structure $\chi_j$ gradually decouple
as half-filling is approached.  We have shown (Section \ref{sec_1el}) that the $n_j$ and $\chi_j$
profiles are locked together (proportional to each other) for $N_c=1$.  Figure \ref{fig:3el_9el} shows
$n_j$ and $\chi_j$ to be very similar for $L{\gg}N_c=3$, while at half filling with $N_c=L=9$ they bear
little resemblance to each other.  In the half-filled situation at large sizes and small $J$, $n_j$ is
essentially constant, while the spatial profile of $\chi_j$ characterizes the so-called Kondo cloud.
\cite{Goth_Assaad, ishii, BarzykinAffleck_PRB98, BordaGarstKroha_PRB09,
  Borda_PRB07, Holzner-etal_PRB09}

\subsection{\label{sec_pert} Analytic expressions for the A region, $J\ll{1}$}

For the A region, one can perturbatively calculate the energy gain $\Delta{E}(J)$.  This gives the
coefficient of the linear ${\sim}L^{-1}J$ dependence in this region.  Also, by examining the
perturbative expressions for $n_j$ and $\chi_j$ profiles, one can expain the close relationship
(Figure \ref{fig:real_space}) between profiles of $N_c=4n\pm1$ pairs, where $n$ is an integer.

\subsubsection{Perturbative calculation for energy gain}

The ground state of the hopping part of the Hamiltonian, is degenerate.  The degeneracy is 2-fold
for $N_c=1$ and 4-fold for $N_c\geq3$.

We first consider $N_c\geq3$.  For $N_c=4n+1$, the four degenerate states for $J=0$ are
\begin{align}
\ket{\Psi_1}=&\ket{\downarrow}\otimes \ket{0 \uparrow,0 \downarrow, \ldots, \mathrm{k_F} \uparrow}, \\
\ket{\Psi_2}=&\ket{\downarrow}\otimes \ket{0 \uparrow,0 \downarrow, \ldots, -\mathrm{k_F}  \uparrow},\notag\\
\ket{\Psi_3}=&\ket{\uparrow}\otimes \ket{0 \uparrow,0 \downarrow, \ldots, \mathrm{k_F}  \downarrow},\notag\\
\ket{\Psi_4}=&\ket{\uparrow}\otimes \ket{0 \uparrow,0 \downarrow, \ldots, -\mathrm{k_F}  \downarrow},\notag
\end{align}
where the states are written as products of the impurity spin state and the bath state.  The
$N_c$-particle bath states are written by specifying the single-particle momentum and spin of the
particles; the states above are obtained by filling up the single-particle states up to the Fermi
momentum.  There are four ground states because there are four single-particle states with moemntum
$\pm{k_F}$, and the last bath electron can fill any one of them.  For  $N_c=4n-1$, the last bath
electron can keep any one of the last four states empty, so that there are also 4 ground states.

The Kondo-coupling part of the Hamiltonian, $H_K$, is now the perturbation.  The matrix elements
$H_{ij} = H_{ji} = \bra{\Psi_i}H_K\ket{\Psi_j}$ of the degeneracy matrix  are found to be
\begin{eqnarray}
H_{ii} = H_{12} = H_{34} =& -\frac{J}{4L} \,,\\
H_{31} = H_{41} = H_{32} =  H_{42}=& \frac{J}{2L} \,.
\end{eqnarray}
The smallest eigenvalue of the $H_{ij}$ matrix gives the most negative energy correction, and hence the
ground-state energy gain at first order is
\begin{equation}\label{eq:energy_gain_pert}
\Delta E = \frac{3}{2}\frac{J}{L} \qquad \mathrm{for} \quad N_c\geq3  .
\end{equation}
The corresponding eigenstate is
\begin{equation} \label{eq:gs_pert}
\ket{\Psi_g} = \frac{1}{2}\left( \ket{\Psi_1}+\ket{\Psi_2}-\ket{\Psi_3}-\ket{\Psi_4} \right) .
\end{equation}

For $N_c=1$, the ground state is 2-fold degenerate: the states can be taken as $\ket{\Psi_1}$ and
$\ket{\Psi_3}$, using the notation introduced above, with $\mathrm{k_F}=0$.  Matrix elements are the
same: $H_{11}=H_{33}=-\frac{J}{4L}$, $H_{31}=\frac{J}{2L}$.  The smallest eigenvalue of the
degeneracy matrix is
\begin{equation}\label{eq:energy_gain_pert_1el}
\Delta E = \frac{3}{4}\frac{J}{L} \qquad \mathrm{for} \quad N_c=1  .
\end{equation}

\subsubsection{Density and spin correlation function at $J\rightarrow0^+$}

Using the ground state \eqref{eq:gs_pert} from perturbation theory we also calculate the electron
density $n_j$ and the spin-spin correlation function $\chi_j$.  After expressing the operators
$\hat{n}_j=\sum_{s}c^\dagger_{j,s}c_{j,s}$ and  $\vec{S}_{\mathrm{imp}}\cdot \vec{s_j} =
\vec{S}_{\mathrm{imp}}\cdot \sum_{s,s'}c_{j,s}^{\dagger}\vec{\sigma}_{ss'}c_{j,s'}$ in momentum
basis, the expectation values in state  \eqref{eq:gs_pert} are found to be
\begin{equation} \label{eq:n_pert}
n_j ~=~ \frac{N_c}{L} - \frac{1}{L} \sin\left(\tfrac{\pi}{2}N_c\right)\cos\left(2\mathrm{k_F} j\right).
\end{equation}
and
\begin{equation}  \label{eq:chi_pert}
\chi_j = -\frac{3}{4L}\left(1+\cos(2\mathrm{k_F}j)\right)
\end{equation}
The factor $\sin\left(\tfrac{\pi}{2}N_c\right)$ used above has values $\pm1$ for $N_c=4n\pm1$;
Eq.\ \eqref{eq:n_pert} explains Figure \ref{fig:real_space}(a).  Eq,\ \eqref{eq:chi_pert} shows that
the impurity-bath spin-spin correlator profiles for  $N_c=4n\pm1$ cases  (same $k_F$) are identical
at small $J$, Figure \ref{fig:real_space}(d).


\section{\label{sec_2D_3D} Higher-dimensional baths}

While our main focus in this paper has been on 1D conduction baths, we will comment now on
higher-dimensional baths. Little changes occur in the regions A and C: Region C is dominated by local
singlet formation, such that $\Delta E\approx \frac{3}{4}J$ continues to hold. Region A remains
perturbative, with $\Delta E\propto \frac{J}{L}$ and a prefactor depending on $N_c$ and the lattice
geometry.

In contrast, changes are expected in the nonperturbative B region, as the phase space (or density of
states) near the band bottom depends on the number of space dimensions. Therefore, the region-B
expression for $\Delta{E}$ in the ultralow-density case (Section \ref{sec_ultralow}) as well as the
region-B$_2$ expression for $\TK$ in the low-density case (Section \ref{sec_RG}) are specific to 1D and
are different in higher dimensions.

As an example, let us consider the ultralow-density case with a single electron $N_c=1$ in an infinite
2D square-lattice bath. The integral expression for the ground state energy $E_0$ in Section
\ref{sec_1el} is modified to
\begin{equation}
\frac{4}{3J} =
-\int\frac{dk_x dk_y /(2\pi)^2}{2\cos\left(k_x\right)+2\cos\left(k_y\right)+E_0} \, .
\end{equation}
Unlike the 1D case, this cannot be solved analytically for $E_0(J)$.  However, for small $J$, the
energy gain $\Delta{E}=-4-E_0$ is found to have the behavior
\begin{equation}
\lim_{J\to 0} \Delta{E} = 32e^{-16\pi/3J} \, .
\end{equation}
Our results in 1D suggest that the energy gain might have the same behavior for any finite odd
$N_c>1$ in an infinite lattice; however, it is difficult to check this conjecture numerically in 2D.

For the low-density (finite $N_c/L$) case, the calculation of Section \ref{sec_RG} is modified in
larger dimensions. For $d=2$ the density of states now follows Eq.~\eqref{dos1} with exponent $r=0$,
such that the tree-level terms in Eq.~\eqref{poor} are absent. Then, the one-loop result for $\TK$ at
$\mu=0$ is\cite{zawa_asy}
\begin{equation}
\label{tkd2a}
\TK \propto D \exp\left(-\frac{3}{4\rho_0 J }\right)
\end{equation}
where the factor $3/4$ is a result of the particle--hole asymmetry; this result continues to hold in
region B$_2$. In contrast, region B$_1$ yields the standard formula
\begin{equation}
\label{tkd2b}
\TK \propto D \exp\left(-\frac{1}{\rho_0 J }\right).
\end{equation}
The crossover between the two regions remains defined by $\TK\sim\mu$ which now gives an estimate for
the boundary as $\rho_0 J_{B12} \sim -1/\ln n_c$.

The behavior in $d=3$ is more intriguing: Here, the density of states is given by Eq.~\eqref{dos1} with
exponent $r=1/2$. As a result of the vanishing $\rho(\omega=0)$ there will be no Kondo screening for
$\mu=0$ and small $J$.\cite{withoff} Instead, a quantum phase transition will occur at $\mu=0$ between
an unscreened and a screened phase upon increasing $J$.\cite{grand_note} This quantum phase transition
will be smeared for $\mu>0$, in a manner similar to that discussed for doped graphene in
Ref.~\onlinecite{epl}. From $\rho(\mu)\propto \mu^{1/2}$ and $n_c\propto\mu^{3/2}$ we can estimate
$\TK$ for small $J$ from the standard Kondo formula, resulting in
\begin{equation}
\TK \propto D \exp\left(-\frac{c_3}{J}\right)
\end{equation}
where $c_3 \propto 1 / (\rho_0 n_c^{1/3})$. This estimate applies to both B$_1$ and B$_2$ regions,
however, with different $c_3$ prefactors in the exponential, similar to Eqs.~(\ref{tkd2a},\ref{tkd2b}).
A more detailed analysis of the crossovers in the $d=3$ case will be given elsewhere.
%


\section{\label{sec_concl} Summary and Conclusions}

In this work we have presented a study of Kondo physics at very low bath densities, so that
particle-hole symmetry is strongly violated.  We have distinguished between two ways of taking the
infinite bath limit, one of them corresponding to the usual thermodynamic limit at low densities,
and the other with fixed particle number, which we call the ultralow-density limit.  To the best of
our knowledge, these regimes have not been studied in depth in the literature (see however
Refs.\ \onlinecite{Zaanen_PRB92, Shchadilova_Ribeiro_arXiv13}).  Either of these two situations may
conceivably be realized in the near future in novel experimental conditions.

In both low-density and ultralow-density cases, there are three clearly distinguishable regions of
coupling: a large $J$ region where one of the bath electrons is strongly localized forms a localized
singlet with the impurity, an intermediate non-perturbative region where the singlet formation is
not sharply localized, and (for finite-sized baths) a small-$J$ perturbative region.  In the
low-density case, the intermediate region is farther demarcated into $J$ values for which the
physics is dominated by the Fermi surface or by the band edge.  Our focus has mostly been specific
to 1D baths, but we have provided some considerations on higher-dimensional baths in the penultimate
section.

The present work opens up several open questions and directions of study.  Perhaps most prominently,
it motivates a full analysis of (ultra)low-density situations in higher dimensions, and also for
various lattice geometries.  Our brief treatment of 2D shows, for example, that the intermediate-$J$
energy gain is described by different nonperturbative behaviors in 1D and 2D.  The role of bath
dimensionality or geometry in determining density profiles ($n_j$ and $\chi_j$) is also currently
unclear.  Other questions include the influence of interactions in the bath --- this has been
addressed for the usual finite-density situation, \cite{Kondo-in-correlated-els_various,
  Kondo-in-correlated_1D} but the effects will conceivably be different in the (ultra)low-density
cases.


\acknowledgments

We thank R. Bulla, L. Fritz, A.~K.~Mitchell, and P.~Ribeiro for helpful discussions and
collaborations on related topics.
This research has been supported by the DFG through FG 960 and GRK 1621 (MV) and by GIF through
grant G 1035-36.14/2009 (MV).



\begin{thebibliography}{99}
\bibitem{grand_note} At $\mu=0$ it is important to work in a grand-canonical ensemble: For large $J$
    there will be one conduction electron in the system which is bound to the impurity.
\end{thebibliography}
\end{document}